\documentstyle[12pt,epsfig]{article}
\newcommand{\be}{\begin{equation}}
\newcommand{\ee}{\end{equation}}
\newcommand{\ba}{\begin{eqnarray}}
\newcommand{\ea}{\end{eqnarray}}
\begin{document}
\title{ Wave Equations With Energy Dependent Potentials.}

\author{ J. Form\'{a}nek\\
Faculty of Mathematics and Physics, Charles University\\
V Hole\v{s}ovi\v{c}k\'{a}ch 2\\
18000 Prague, Czech Republic\\
R.J. Lombard\\
Groupe de Physique Th\'eorique \\
Institut de Physique Nucl\'eaire\\
91406 Orsay Cedex, France\\
and J. Mare\v{s}\\
Nuclear Physics Institute\\
250 68 \v{R}e\v{z}, Czech Republic}

\maketitle

Abstract :

We study wave equations with energy dependent potentials. Simple
analytical models are found useful to illustrate difficulties
encountered with the calculation and interpretation of
observables. A formal analysis shows under which conditions such
equations can be handled as evolution equation of quantum theory with
an energy dependent potential. Once these conditions are met, such theory
can be transformed into ordinary quantum theory.

\newpage

\section{Introduction}

Wave equations with energy dependent potentials have been known
for long time. They appeared along with the Klein-Gordon equation
for a particle in an external electromagnetic field~\cite{SSW}. In
non-relativistic quantum mechanics, they arise from momentum
dependent interactions, as shown by Green \cite{AMG}. The
Pauli-Schr\"{o}dinger equation represents another
example~\cite{WPA,BSA}. A Hilbert space formulation of the
relativistic quantum mechanical two-body problem was studied by
Rizov {\it et al} \cite{RST}. These authors derived an algorithm
for a perturbative calculation of the energy eigenvalues in the
case of a general energy dependent quasipotential. The problem was
first solved for the particular case of the relativistic harmonic
oscillator \cite{LST}.

In recent years, numerous works have been devoted to the
Hamiltonian formulation of relativistic quantum mechanics in
connection with the manifestly covariant formalism with
constraints \cite{HS1,CVA}. The main advantage of this approach is
to allow the proper separation of relative and center-of-mass
coordinates in few-body and even in many-body problem, and thus to
eliminate spurious degrees of freedom. The constraints impose
restrictions to the potentials to be physically acceptable. As a
result, the potentials become energy dependent. Links with field
theory have been developed by taking the Bethe-Salpeter equation
as a starting point. This makes the approach a challenging one. A
typical application of the manifestly covariant formalism is a
calculation of spectra of baryonic atoms. Work along this line has
been carried out by Mourad and Sazdjian \cite{MHS}. In particular,
they have derived two-fermion relativistic wave equations, which
acquire the form of Pauli-Schr\"odinger equations.

The presence of the energy dependent potential in a wave equation
has several non-trivial implications. The most obvious one is the
modification of the scalar product, necessary to ensure the
conservation of the norm \cite{HS2}. However, the modification of
the scalar product itself is not sufficient to justify the
use of the common rules of quantum mechanics. In the next
section, we illustrate the aforementioned difficulties on a few
simple models, admitting analytical solution.

Of course, testing formulas on a particular model does not guarantee
their general validity in more realistic, physical systems.
In section 3, we therefore analyze formal aspects and demonstrate
under what circumstances a wave equation with an energy dependent
potential corresponds to an acceptable quantum theory. We show
that in such a case the wave equation can be transformed into an
equivalent ordinary Schr\"odinger equation with a local, momentum
dependent potential.
Implications of the formal treatment are illustrated in section 4 and
 conclusions are drawn in the last section.

\section{A toy model.}
\subsection{Generalities.}

For the sake of simplicity, the toy model is constructed in the 1-dimensional
space.\footnote{
Throughout this section, we assume that solving the wave equation with an
energy dependent potential we are able to select the states in such a way
that each eigenvalue corresponds to a single normalizable wave function.
This is not guarantee {\it a priori}, as will be discussed in the next
section.}

The starting point is the time dependent wave equation\footnote
{In this work we use $m$ = $\hbar$ = 1.}
\be
i \frac{\partial}{\partial t} \Psi(x,t) = [ - \frac{1}{2}\frac{\partial^2}{\partial x^2}
+ V(x,i\frac{\partial}{\partial t}) ] \Psi(x,t) \ ,\ee
where $V(x,i\frac{\partial}{\partial t})$ denotes a real function of 2
variables.
Setting $\Psi(x,t) = e^{-iEt}\psi(x) $ yields
\be
H\psi(x) = [ - \frac{1}{2}\frac{\partial^2}{\partial x^2}
+ V(x,E) ] \psi(x) = E\psi(x)  \ . \ee

The first modification of the usual rules of quantum mechanics concerns the
scalar product. This question has been studied by many authors in the one- and
two-particle relativistic quantum mechanics (see
Sazdjian \cite{HS1} and references therein). For completeness,
we present a brief sketch  in the 1-dimensional case :

Consider two solutions of energies $E$ and $E'$
\be
\Phi_{\epsilon}(x,t) = e^{-i(E -i\epsilon)t}\Phi(x) ,\ \ \  {\rm and}
\ \ \ \ \Psi_{\epsilon}(x,t) = e^{-i(E' -i\epsilon)t}\Psi(x) , \ee
with $\epsilon \rightarrow 0$.
The usual continuity equation reads
\be
\frac{\partial}{\partial t}P  =  - \frac{\partial}{\partial x} j \ , \ee
where
$$
P = \Psi^*_{\epsilon}(x,t)   \Phi_{\epsilon}(x,t) \ \ ; \ \ j =
- i \Big[\Psi^*_{\epsilon}(x,t) (\frac{\partial}{\partial x}  \Phi_{\epsilon}(x,t))
-(\frac{\partial}{\partial x}\Psi^*_{\epsilon}(x,t) )  \Phi_{\epsilon}(x,t) \Big]  \ .
$$

In the case of energy dependent potential, in order to fulfill the
continuity equation,
 a term
\be
\frac{\partial}{\partial t} P_a = i \Psi_{\epsilon}^*(x,t)[ V(x,E) - V(x,E') ]
 \Phi_{\epsilon}(x,t) \  \ee
has to be added to the left hand side of Eq. (4).
 After integration, Eq. (5) yields
\be
P_a = - \Psi_{\epsilon}^*(x,t) \Big[ \frac{V(x,E') -V(x,E)}{E' - E} \Big]
\Phi_{\epsilon}(x,t) \ , \ee
as $\epsilon \rightarrow 0$. As a result, in the limit
$E' \rightarrow E$, the scalar product (the norm) is given by
\be
N = \int_{-\infty}^{\infty} \Psi^*(x)[1 - \frac{\partial V}{\partial E}
]\Psi(x) dx \ . \ee
Let us specify the stationary states by a global quantum number $n$.
The orthogonality relation between
two states $n$ and $n'$, $n \neq n'$, is given by
\be
\int \Psi_{n'}^*(x)
[1 - \frac{V(x,E_{n'}) - V(x,E_n)}{ E_{n'} - E_n}] \Psi_n(x) dx =
0 \  \ . \ee
The generalization to higher space dimensions is straightforward.
Expectation values and transition matrix elements have to be calculated in a
similar way, i.e. by including the correction term.
Use has been made here of the continuity equation. The form of the scalar
product with energy dependent potentials was also  derived from Green's
functions of quantum field theory~[11].

The modification of the scalar product is a necessary but not sufficient
step. A coherent theory requires the norm and the average values of positive
definite operators to be positive. This is not guaranteed {\it a priori}.

The second difficulty to face is the loss of the usual completeness relation
$\sum_n \Psi_n(x')\Psi_n^*(x) = \delta(x - x')$.
Of course, this is a direct consequence of the fact that the
functions $\Psi_n(x)$ constructed by the described procedure do not
represent eigenfunctions of the same (linear selfadjoint) operator
on $L^2(-\infty,\infty)$.
It is tempting to propose :
\be
\sum_n \Psi_n(x')\Psi_n^*(x)[1 - \varphi_{nn}(x)] = \delta(x - x')
\ ,
\ee
$$
 {\rm where  } \ \ \ \varphi_{n n'}(x) =
\frac{V(x,E_{n'}) - V(x,E_n)}{ E_{n'} - E_n}
\ \ {\rm for\ n \neq n'} , \ \ {\rm and  } \ \ \  \varphi_{n n}(x) = \frac{\partial V(x,E_n)}{\partial E} \ . $$
According to this ansatz, we have
\ba
\int \Psi_j(x) \delta(x - x')dx  & = & \Psi_j(x') \\ \nonumber
  &  = &
\sum_n \ \Psi_n(x') \int \Psi_j(x)
\Psi^*_n(x)[1 - \varphi_{nn}(x)] dx \\ \nonumber
&    = & \Psi_j(x')  - \sum_n \Psi_n(x') \int \Psi_j(x)
  \Psi^*_n(x) [ \varphi_{nn}(x) - \varphi_{nj}(x) ] dx \ . \ea
The second term is non zero in general. Thus, the ansatz (9) is not valid except
for the case of a linear energy dependence, when the terms
$\varphi(x)_{nj} = \varphi(x)$ are
independent of the state.

The third complication arises due to the presence of the eigenvalue in the
Hamiltonian (2). When calculating a commutator, one has to replace the eigenvalue by the
corresponding operator.
For the sake of illustration, let
us take the following Hamiltonian :
\be
H = \frac{p^2}{2} + V_0(x) + V_1(x)E \ .\ee
The commutator $\Big[H,x\Big]$ is given by
\be
\Big[H,x\Big] = -i p + V_1(x)\Big[H,x\Big] =
 -i \Big(1 - V_1(x) \Big)^{-1}p  \ . \ee
It is to be stress here that $p=-i\frac{\partial}{\partial x}$
not necessarily represents the momentum operator.

Similarly, the double commutator takes the form
\be
\bigg[\Big[H,x\Big],x\bigg] = -  \Big(1 - V_1(x)\Big)^{-1} \ . \ee
In general, the commutators are more difficult to calculate, but the
procedure is the same. It starts with
\be
\Big[H,x\Big] = -ip + \frac{\partial V(H,x)}{\partial H}
\frac{\partial H}{\partial p} \ . \ee

\subsection{Explicit examples.}

In order to construct a toy model admitting analytical solutions, we consider
the harmonic oscillator potential with an energy dependent frequency.
In 1 dimension, the corresponding wave equation reads
($\omega$ = 1)
\be
\Big [ - \frac{1}{2}\frac{d^2}{dx^2}  +  \frac{1}{2}(1 + \gamma
f(E_n))x^2 \ \Big ]\Psi_n(x) = E_n \Psi_n(x) \ , \ee
where $\gamma$ is  a parameter (not necessarily small though most of our
investigations are dedicated to small $\gamma$).
Setting
\be
\lambda_n^2 = 1 + \gamma f(E_n) \ \  , \ \
k_n^2 = 2E_n \ , \ee
leads to
\be
\Big [ \frac{d^2}{dx^2} + k_n^2 - \lambda_n^2 x^2 \Big] \Psi_n(x) = 0.\ee
The solutions are obtained by the ansatz
\be
\Psi_n(x) = C_nh_n(\sqrt{\lambda_n}x) e^{-\lambda_n x^2/2} \ , \ee
which yields for $\sqrt{\lambda_n}x = z$
\be
\frac{d^2 }{dz^2}h_n(z) - 2 z \frac{d}{dz}h_n(z) +
(\frac{k_n^2}{\lambda_n} - 1) h_n(z) = 0 \ . \ee
Apart from the fact that $\lambda_n$ depends on the state,
the solutions of this equation are formally the same as the Hermite
polynomials.
Their orthogonality is ensured by the weight function
\be
e^{-(\lambda_n + \lambda_{n'}) x^2/2} [1 -
\frac{\gamma}{2} \ \frac{f(E_{n'}) - f(E_n)}{(E_{n'} - E_n)}
\ x^2] \ .\ee
The energies and the $\lambda_n$'s are obtained from Eq. (16) together with
\be
k_n^2 = \lambda_n(2n + 1) = 2 E_n \ . \ee
Two kinds of energy dependence have been considered : the linear and the
square root cases.

\subsubsection{Linear E-dependence}

For $f(E_n) = E_n$,
$\frac{\partial V(x,E_n)}{\partial E_n} = \frac{\gamma}{2} x^2$
and the conditions to be satisfied lead to a quadratic
equation. Its general solution is given by
\be
E_n = \frac{(2n+1 )^2}{8} \gamma  \pm (n + \frac{1}{2})
 \ \
 \sqrt{ 1 + \frac{(2n + 1)^2}{16}  \gamma^2} \ .\ee
The solutions have two branches, one of positive and one of negative values.
However, only the positive energies are
retained, since the negative ones are not normalizable.

In Fig.~1, the energy $E_n$ is presented as function of n for several values of
$\gamma$ and $n \leq 8$.
While positive values of $\gamma$ produce an increase of the level spacing
as function of $n$, negative values of $\gamma$ compress the spectrum.
For $\gamma < 0 $,
$E_n$ is a monotonically increasing function of $n$. Asymptotically
($n \rightarrow \infty$), its derivative reaches
$-\frac{24}{(2n+1)^3 \gamma^3}$, and $ E_n \approx -\frac{1}{\gamma}$.
This limit ensures that $\lambda_n$ is
 real   $ \forall n$, with $\lambda_n \rightarrow 0$.

As stated in the preceding section, physically acceptable potentials must
yield positive norms. In the present case, we have
\be
C_n^2 \int_{-\infty}^{\infty} e^{-\lambda_n x^2}
h_n^2(\sqrt{\lambda_n}x) [ 1 - \frac{\gamma}{2} \  x^2 ] dx = 1 \ , \ee
which gives
\be
C_n^2 \ = \  \frac{1}{2^n \ n!} \
\frac{\sqrt{\lambda_n}}{\sqrt{\pi}} \Big[ 1 -
\frac{\gamma}{4 \lambda_n} (2n + 1)  \Big ]^{-1}\ .
\ee
From Eqs. (16) and (21), we get
\be
\frac{(2n + 1)}{4}\gamma = \lambda_n -
\sqrt{ 1 + \frac{(2n + 1)^2}{16} \gamma^2} \ . \ee
Consequently,  the norm is positive definite $ \forall n$, irrespective of
the sign of $\gamma$.

The modification of the scalar product affects also the calculation of
expectation values. In the case of the mean square (ms) radius
we obtain
\be
\langle x^2 \rangle_n  =
\frac{(2n + 1)}{2 \lambda_n} \Big[ 1 - \frac{\gamma}{4 \lambda_n}
(2n + 1)  \Big]^{-1} \ \Big[ 1 - \frac{3 \gamma}{8 \lambda_n} \frac{[(2n
+ 1)^2 + 1]}{(2n + 1)}   \Big]\ . \ee
For $\gamma$ positive,
the sign of $\langle x^2 \rangle_n$ is linked to the sign
of the last term. It is easy to show that for $n \rightarrow \infty$ it takes
the value 1/4, whereas for $n=0$, it becomes negative beyond
$\gamma = \sqrt{16/3}$. Thus the question is, for which state
the ms radius is positive definite. For $n=1$, it is
straightforward to show (using Eqs. (16), (21)) that the last factor of
Eq. (26) has no real root,
and the ms radius is  thus positive definite.

As the critical value $\gamma = \sqrt{16/3}$ is large compared to 1, one
might expect to avoid contradictions for $\gamma << 1$. Unfortunately,
explicit calculations of higher moments $\langle x^k \rangle_n$ ($k$ even)
for few low $n$ states indicate
that (up to the leading order):
\be
\langle x^k \rangle_n \propto
 [1 - \frac{\gamma \ (k + 1)}{4 \lambda_n} ] \ . \ee
Thus, for fixed $n$ and positive $\gamma$, there will always be a value
of $k$, beyond which $\langle x^k \rangle_n$
becomes negative.

The above example clearly demonstrates that ad-hoc modification of the
operation which should play the role of the
scalar product and the positivity of the ``norm'' of the assumed energy
eigenfunctions are not sufficient to ensure
the positivity of the expectation value of a positive definite operator.
It underlines an internal inconsistency for $\gamma > 0$.

As far as the closure relation is concerned, as stated in the preceding
section, the ansatz (9) seems to be valid for a linear energy dependence.
It is straightforward to get a numerical test by calculating, for instance,
\be
\langle x^2\rangle_0 = \sum_n|\langle n|x|0 \rangle|^2 \ . \ee
Up to linear terms in $\gamma$, we verified that
\be
\langle x^2 \rangle_0 \approx   \frac{1}{2}  ( 1 - \frac{3}{4} \gamma )
 \approx  |\langle 1|x|0 \rangle|^2  \ . \ee
Higher contributions,  $|\langle n|x|0 \rangle|^2 \ , \ \
{\rm where} \ \ n \geq 3 $,
are proportional at least to $\gamma^2$.
This results from the fact that in the ordinary harmonic oscillator
 ($\gamma = 0$), the $x$
operator connects only states which differ by one quantum of energy.
It is obviously a tedious work to check Eq. (28) to all orders in $\gamma$.
However, since we know the exact value of
$\langle x^2 \rangle_0$, we can estimate the
contributions from the first excited states and discuss the convergence of
the series expansion. As an example, we have calculated the first two
terms for $0.01 \leq |\gamma| \leq 0.50$. The results are displayed in Table 1.
They are compared with the exact values and with the estimates obtained
keeping only contributions linear in $\gamma$. Whereas the latter approximation
becomes inaccurate by few percent beyond $\gamma = 0.25$, the convergence seems
rapid.
The contribution of the third excited state is less than 2 \% for $\gamma = 0.5$.

As the second example, we have tested the same way
\be
\langle x^4 \rangle_0 = \sum_n |<0|x^2|n>|^2
\approx |\langle 0|x^2|0 \rangle|^2 +
|\langle 2|x^2|0 \rangle|^2 \approx \frac{3}{4}
( 1 - \frac{3}{2} \gamma ) \ , \ee
keeping only terms linear in $\gamma$.
In Table 2, we compare the exact values with the linear estimates
and with the sums of the series expansion as a function of $n_{max}$
 (the highest state involved in the sum).
The situation is somewhat worse than for $<x^2>_0$ :
the linear approximation is less
efficient and the convergence of the series expansion is slower.

The above results support the extension (9) of the closure relation.
They also underline the care one has to take when using
$\gamma$ as a small parameter in a series expansion.
On the other hand, for $\gamma > 0$, we note a tendency to overpass the
exact value, which appears beyond $\gamma = 0.4$ for $<x^2>_0$ and beyond
$\gamma = 0.35$ for $<x^4>_0$. We have verified that this effect is not due
to numerical uncertainties. Again, it is a signature of the deficiency of the
algorithm for $\gamma >0$.

The consequence of the modification of
commutation relations can be illustrated on
the dipole sum rule.
We recall that it is obtained by
averaging over the ground state the double commutator
$\Big [ [H,x],x \Big]$.

According to Eq. (13), we have
\be
\Big [ [H,x],x \Big]  = - \frac{1}{1 - \gamma x^2/2}
\ . \ee
and the ground state expectation value yields
\be
\sum_n(E_n - E_0) |\langle n|x|0 \rangle|^2 = \frac{1}{2}
\langle 0|\Big [ [H,x],x \Big]|0 \rangle =
\frac{1}{2} [1 - \frac{\gamma}{4\lambda_0}]^{-1}  \ .\ee
As a numerical test, we adopted several maximum values $n_{max}$
of $n$ considered in the sum. Results are
displayed in Table 3 for $n_{max} = 1$ up to 7. They are compared with the
exact value given by the last term of Eq. (32). The convergence rate is
slower as $\gamma$ increases, as expected. The convergence towards the right
result for the low $\gamma$ values supports the validity of the rules used
to derive Eq. (32). For $\gamma >0$, contrary to the  $<x^2>_0$ and
$<x^4>_0$ cases, the sum seems to converge toward the exact value. This comes
from the fact that the leading contribution to the ground state average of
Eq. (32) is the norm, for which the orthonormality of the wave functions
ensures that $\sum_n |<n|0>|^2 = 1$.

\subsubsection{$\sqrt{E}$ - dependence}

The $\sqrt{E_n}$ case is chosen for the sake of comparison with the
results of the previous example. A smoother energy dependence should generate
weaker effects. On the other hand, the simpler generalization (9) of the
closure relation is valid only approximately. Therefore, we do not expect to
calculate expectation values of commutators (13,14) with unlimited accuracy.

The eigenvalues of Eq. (15) are in this case  given by
\be
4 E_n^2 - (2n +1)^2 [1 + \gamma \sqrt{E_n} ] = 0 \ .\ee
On the real half-plane $E_n \geq 0$,
for each sign of $\gamma$ and each $n$,
there exists a single real eigenvalue leading to
square integrable wave function. The spectrum exhibits the same features as
the linear $E_n$-dependence.

For small positive $\gamma$ (typically
$\gamma \leq 0.2$) and few lowest states ($n \leq 3$),
a perturbative estimate
yields
\be
E_n \approx (n + \frac{1}{2} ) [1 + \frac{\gamma}{2}
\sqrt{\frac{(2n+1)}{2}} ]\ .\ee

For $\gamma < 0$, the large $n$ limit is given by
\be
\lim_{n\to\infty} E_n = \frac{1}{\gamma^2} \ . \ee
As in the linear $E_n$-dependence, it ensures $\lambda_n$ to be real and
reaching 0 as $n \rightarrow \infty$.

The norms and the (assumed) predictions of the average values of
observables are calculated by introducing the multiplicative
factor
\be
1 -
\frac{\gamma}{2(\sqrt{E_n} + \sqrt{E_{n'}})}x^2 \ . \ee

The norms, given by
\be
C_n^2 = \frac{1}{2^n n!}\frac{\sqrt{\lambda_n}}{\sqrt{\pi}}
\Big [ 1 - \frac{1}{8 \lambda_n} \gamma \frac{(2n + 1)}{\sqrt{E_n}}
\Big ]^{-1} ,\ee
are positive definite as can be proved using Eq. (21) and (33).

The ms radius
\be
\langle x^2 \rangle_n = \frac{2n + 1}{2 \lambda_n} \Big [ 1 - \frac{1}{8
\sqrt{E_n}} \frac{\gamma (2n + 1)}{\lambda_n} \Big]^{-1}
\Big[ 1 - \frac{3}{16 \sqrt{E_n}} \
\frac{\gamma [(2n + 1)^2 + 1]}{\lambda_n (2n + 1)}
\Big]\  \ee
 is positive definite $\forall n$, as well.
On the other hand, for $\langle x^k \rangle_n$, the same reasoning as for
the linear case yields (up to the leading order)
\be
\langle x^k \rangle_n \propto [1 - \frac{\gamma (K + 1)}{4 \lambda_n
\sqrt{E_n}} ] \ . \ee
For fixed $\gamma > 0$  and $n$, there exists a value of $k$
beyond which $\langle x^k \rangle_n$ becomes negative.
In spite of a smoother energy dependence
than in the linear case, very similar problems occur in the case of
expectation values of positive definite operators for $\gamma>0$.

Because $\varphi_{n n'}$ are energy dependent, the "closure ansatz" has no chance
to be given by the simple extension~(9). Nevertheless, one can estimate the
correction terms. To fix the order of magnitude, we take Eq. (10) for
$\Psi_j = \Psi_0$,
where
\be
[ \varphi_{nn}(x) - \varphi_{n0}(x) ] =
\frac{\gamma \ x^2}{2}\ [\ \frac{1}{2\sqrt{E_n}} - \frac{1}{(\sqrt{E_0} +
\sqrt{E_n})} \ ] \  . \ee
We consider the first two contributions to the sum of Eq (10).
Because of parity, they arise from
$\Psi_2$ and
$\Psi_4$. The
results are displayed in Table 4. The correction is small in the range of
considered $\gamma$ values, but not necessary negligible. It is expected to
decrease with increasing $n$.

Neglecting the correction terms, we can perform the same tests as in the
preceding subsection. Up to linear terms in
$\gamma$,
$$
\langle x^2 \rangle_0 \approx \frac{1}{2}(1 - \frac{\gamma}{\sqrt{2}}) \ \
{\rm is \ to \ be \ compared \ to } \ \
|\langle 1|x| 0 \rangle |^2 \approx \frac{1}{2}
(1 - 0.65 \gamma )  \ . $$
The difference between the two estimates is of the same order as the
correction terms listed in Table 4.

A more quantitative estimate is displayed in Table 5. The
exact value of $\langle x^2 \rangle_0$ is compared with the sums
$\sum_n^{n_{max}}|\langle n|x| 0 \rangle|^2$, with $n_{max}$ = 1,2 and 3. The
series expansion converges rather rapidly but not to the exact value. The
discrepancies are of the same order as the one observed in the above
expression limited to terms linear in $\gamma$.

\subsubsection{Further energy dependences}

The above two examples yield an overview of the difficulties linked to the
energy dependence of the potential in a wave equation. There is no need to
pursue this investigation much further. One point, however, is worth
mentioning. It concerns the cases
when higher powers of the energy dependence are
considered. Take for instance  $f(E_n) = E_n^2$. It is easy to verify that
the positive eigenvalues are given by
\be
E_n =  \frac{2n + 1}{\sqrt{4 - (2n + 1)^2\gamma}} \ . \ee
We note that for positive $\gamma$ there will always be a critical $n$
beyond which the eigenvalue becomes complex.

\section{ A quantum theory.}

A toy model is well suited for illustrative purposes. However, in order to
go beyond and check if it corresponds to a well defined theory, a more
fundamental analysis is required.
Mathematical aspects of theories with energy dependent potentials have been
studied in ref. [12,13,5]. In particular, these works are devoted to the
positivity of the norm and the occurance of pathologies~[1].

Let us define (for a given Hilbert space $\mathcal{H}$) a Hamiltonian
\be
\hat{H}(z) = \hat{H}_0 + \hat{V}(z) \ , \ee
where $\hat{H}_0$ is a selfadjoint operator and there exists such a set
$\mathcal{M} \subset \mathcal{R}$ that for any fixed value
$z \in \mathcal{M}$ each of the operators $\hat{V}(z)$, $\hat{H}(z)$ is
selfadjoint too.

Let us make attempt to construct a quantum theory (QT) in such a way that

1) the vectors $ |\Psi> \in \mathcal{H}$, at least some of them, correspond
to states of a physical system (here and in the following ``state'' is used
in the sense of ``pure state'' only),

2) the time evolution of these states is determined by the equation
\be
i \frac{d}{dt} |\Psi(t)> = \hat{H}(i\frac{d}{dt})|\Psi(t)> \ . \ee
First of all, one should keep in mind the following requirements, which have
to be satisfied by {\bf any} QT :

i)    Each state of the considered physical system is mapped to a ray in the
corresponding Hilbert space (${\mathcal{H}}_p $).\\
ii)   Each state of the considered system represents a linear combination of
its stationary states.\\
iii) The initial state of the system (its state at $t_0$) uniquely
determines its states at each later moment $t>t_0$ (provided the system was
not submitted to any measurement and/or observation during the interval
($t_0$,$t>$ ).

The last requirement apparently limits the acceptable $z$-dependence of
$\hat{V}(z)$ to the linear one only. But one can try to avoid this sharp
restriction in the following way : the Hilbert space ${\mathcal{H}}_p $ need
{\bf not} be identical with $\mathcal{H}$. In fact, the two spaces might
differ in 2 aspects :

a) some of the vectors from $\mathcal{H}$ need not belong to
${\mathcal{H}}_p $; these elements of $\mathcal{H}$ would not describe any
state of the considered system, i.e., the vector space ${\mathcal{H}}_p $
could be a (proper) subspace of the vector space $\mathcal{H}$.

b) The vector space ${\mathcal{H}}_p $ could be endowed with a different
scalar product ($ \big(\cdot \ | \ \cdot >$ )
than the vector space $\mathcal{H}$. (We
reserve the symbol $<\cdot|\cdot>$ for the scalar product in the Hilbert space
$\mathcal{H}$). In other words, the Hilbert space ${\mathcal{H}}_p $ need
{\bf not} be a subspace of the Hilbert space $\mathcal{H}$.

From the requirement i) directly follows that a vector
\be
|\Psi> \in {\mathcal{H}}_p  \ \   ,\ \  |\Psi> \not= 0 \ee
corresponds to a stationary state of the considered system if and only if
there is such a real number $E$ that the vector valued function (of time)
\be
|\Psi(t)> = e^{-iEt}|\Psi> \ee
satisfies Eq. (43).
Moreover, it is evident that the function (45) solves Eq. (43) if and only
if the vector (44) satisfies the relation
\be
( \hat{H}_0 + \hat{V}(E)\ ) \ |\Psi> = E \  |\Psi>  \ \  .\ee
It is customary to call this relation the (time independent) ``Schr\"odinger
equation with an {\bf energy dependent potential}''.

Let us recall that $\forall E \in \mathcal{M}$ there
$\exists \  |\Psi> \in \mathcal{H}$;
\be
( \hat{H}_0 + \hat{V}(E) \ ) \ |\Psi> \ \in \ \mathcal{H} \ .\ee
On the other hand (usually) for most of values $ E \in \mathcal{M}$ none of
\be
|\Psi> \ \in {\mathcal{H}} \ , \ | \Psi >  \not= 0  \ee

satisfies the condition (46). To find out those values of $E$ for which
Eq. (46) has nontrivial solution in $\mathcal{H}$, let us first of all
consider the equation
\be
\hat{H}(z) |\Psi> = E(z) \ |\Psi> \ . \ee
For simplicity, we will limit our discussion to the case where
$\forall z \in \mathcal{M}$ the spectrum of $\hat{H}(z)$ is purely
{\bf discrete} and {\bf non-degenerate}. This means that for any fixed value
$z \in \mathcal{M}$, to each eigenvalue $E_n(z)$ of $\hat{H}(z)$, there
$\exists \  |\Psi(z)> \ \in \mathcal{H}$ :
\be
\hat{H}(z) |\Psi_n(z)> = E_n(z)\ |\Psi_n(z)> \ , \ \ \  |\Psi_n(z)> \not= 0 \ee
and no other linearly independent solution of Eq. (46) with $E= E_n(z)$ exists
in $\mathcal{H}$. It may happen that for a given value of $n$ there exist
such values $z \in \mathcal{M}$ that the relation
\be
E_n(z) = z \ee
holds. In the following, we will enumerate by the index $n$ {\bf only}
those eigenvalues of $\hat{H}(z)$ for which Eq. (51) has at least one
solution in $\mathcal{M}$. Those values of $z \in \mathcal{M}$ which satisfy
the relation (51) will be denoted as
\be
z_m^{(n)} \ , \ m = 1,\ldots ,M_n \geq 1 \ , \ee
i.e., $z_m^{(n)}$ are  those elements of $\mathcal{M}$ for which
\be
E_n(z_m^{(n)}) = z_m^{(n)}  \ . \ee
Due to Eq. (50) we have
\be
\hat{H}(z_m^{(n)}) \ |\Psi_n(z_m^{(n)})> = E_n(z_m^{(n)}) \
|\Psi_n(z_m^{(n)})> = z_m^{(n)} \
|\Psi_n(z_m^{(n)})> \ .\ee
Moreover, due to the assumed non-degeneracy of the $\hat{H}(z)$ spectrum, we
know that for each given $z_m^{(n)}$ the last equation has one and only one
linearly independent solution in $\mathcal{H}$. Let us denote it as
$|\Psi_m^{(n)}>$. Without any loss of generality, on can impose the
normalization condition
\be <\Psi_m^{(n)} \ | \ \Psi_m^{(n)} > = 1 \ . \ee
In order to simplify symbolics, let us switch from the two-index to one-index
notation
\be
\{(n),m\} \rightarrow j \equiv \sum_{k=0}^{n-1} M_k + m \ , \ \
{ \rm where} \ \
  M_0 = 0 \ee
and
\be
z_j \equiv z_m^{(n)} \ , \ |\Psi_j> \equiv \ |\Psi_m^{(n)}> \ .\ee

If one accepts the ``philosophy of energy dependent potential'', then each
of the values
\be
z_j \ , \ j = 1,\ldots ,N \equiv \sum_k M_k \ee
represents an energy level of the considered system (and each energy level
coincides with one of the values (58)) and the vector $|\Psi_j>$
unambiguously describes the corresponding stationary state.

In order to satisfy the aforementioned condition ii), one has to require
 $\forall |\Psi> \in {\mathcal{H}}_p$ to be expressible in the form
\be
|\Psi> = \sum_j \ c_j \ |\Psi_j> \ , \ \
{\rm where} \ \  c_j = const. \ . \ee

Such a vector space ${\mathcal{H}}_p$ could represent Hilbert space assigned
to a physical system in the framework of {\bf any} QT only if it could be
endowed with such a scalar product $\big( \cdot|\cdot >$ that the
orthogonality condition
\be
\big( \Psi_j | \Psi_k> = 0  \ \ , \ \ \forall j \not= k \  \ee
would be guaranteed.
On the other hand,
\be
(\hat{H}_0 + \hat{V}(z_j) \ | \ \Psi_j> = z_j \ | \Psi_j> \ee
implies  (due to selfadjointness of $\hat{H}_0$ and $\hat{V}(z)$) that the
identity
\be
(z_j - z_k) <\Psi_k | \Psi_j> = < \Psi_k | [\hat{V}(z_j) - \hat{V}(z_k)]
|\Psi_j> \ \ee
is valid  $\forall j \not=k$.

Hence, the requirement (60) would be automatically satisfied if the scalar
product in ${\mathcal{H}}_p$ is defined in such a way that
\be
\big( \Psi_j | \Psi_k> \equiv <\Psi_j | \hat{W}_{jk} | \Psi_k> \ , \ee
where
\be
\hat{W}_{jk} \equiv 1 - \hat{Q}_{jk} \ee
and
\be\hat{Q}_{jk} \equiv \frac{ \hat{V}(z_j) - \hat{V}(z_k)}{ z_j - z_k} \ \
,  \ \ \forall \  j \not= k \ \ . \ee

In order to define the scalar product in ${\mathcal{H}}_p$, it remains to
specify the meaning of the r.h.s. of the expression (65) for $j=k$. It is
tempting to take
\be
\hat{Q}_{jj} \equiv \frac{d\hat{V}}{dz} (z_j) \ . \ee
We are not going to dwell on arguments neither in favor nor against this
specific choice of the ``scalar product'' $\big(\cdot|\cdot>$,
utilized e.g. in Eq. (8).
Let us stress instead that one can face serious
problems when introducing  {\bf any} scalar product $\big(\cdot|\cdot>$,
which would guarantee fulfillment of the requirement (60).
The point is that the conditions (54) and (55)
define the set of vectors
\be
|\Psi_j> \ , \ j=1,\ldots ,N \ee
(up to  phase factors) uniquely but they do not imply the linear independence
of all the vectors (68). One should not forget that the requirements (54)
and (55)
{\bf do not} guarantee that all the vectors (67) are eigenvectors of a
{\bf common} selfadjoint operator. On the other hand, if
\be
|\Psi_j> = \sum_{k \not= j} C_k \ | \Psi_k> \  ,\ee
then from the condition (60) follows
\be
\big( \Psi_j | \Psi_j> = 0 \ , \ee
i.e.,  $|\Psi_j>$ would be the null vector. But from the condition (55)
we know that it is not so.
This means that if not all the vectors (68) are linearly independent, than
it is {\bf impossible} to define {\bf any} scalar product $\big(\cdot|\cdot>$ in
${\mathcal{H}}_p$ in such a way that the condition (60) is satisfied.
Especially, this means that in such a case\\
i)   the relations (63)-(66) {\bf do not} define a {\bf scalar product}\\
ii)  it is {\bf impossible} to modify the r.h.s. of equation (66) in such a
way that the relations (63)-(66) would define a scalar product in
${\mathcal{H}}_p$.\\
In other words, in such a case, the aforementioned algorithm {\bf does not}
represent {\bf any} QT.

On the other hand, it is clear that for any given quantities $\hat{H}_0$,
$\hat{V}(z)$ there exists (at least one) such a set $\mathcal{M}$ that all
the vectors (67) are linearly independent. Therefore, at least in
principle, one can avoid the above indicated problem simply by a
``proper choice'' of the set $\mathcal{M}$.

Of course one should keep in mind that\\
$\alpha$) the requirement of the linear independence of the vectors (67)
{\bf does not} determine the set $\mathcal{M}$ uniquely. ( It could be
tempting to choose $\mathcal{M}$ as the largest subset of real
values $z$ for which the operators $\hat{H}(z)$, $\hat{V}(z)$ are
selfadjoint and all the vectors (67) are linearly independent. But generally
there still could be more then one subset of $\mathcal{R}$ which fulfills all
these conditions.)\\
$\beta$) the ``energy spectrum'' produced by such an algorithm substantially
depends on the \ \ \  particular choice of the set $\mathcal{M}$.

The main moral from these observations is that the specification of the operator
(valued function) (42) itself {\bf need not} determine uniquely the energy
spectrum of the corresponding physical system with energy dependent potential.

To proceed further in our analysis, let us from now on assume that the set
$\mathcal{M}$ is already chosen in such a way that all the vectors (67) are
linearly independent, and therefore the vector space ${\mathcal{H}}_p$
 (59) is a $N$-dimensional subspace of the vector space
$\mathcal{H}$. Let us also define vectors
\be
|\Psi^j> \equiv A^{(j)} \ |\Psi_j> \ , \ee
where $A^{(j)}$ are non-vanishing constants determined (up to their phase
factors) by the requirement
\be
\big( \Psi^j | \Psi^k> = \delta_{jk} \ ; \ j,k =1,\ldots ,N \ , \ee
once the scalar product $\big(\cdot|\cdot>$ in ${\mathcal{H}}_p$ is specified.

Since ${\mathcal{H}}_p$ is an $N$-dimensional subspace of
$\mathcal{H}$, there exist such vectors
\be
| \chi_j> \ \in {\mathcal{H}}_p \ , \ j=1,\ldots ,N
\ee
that
\be
<\chi_j | \chi_k> = \delta_{jk} \ , \ j,k = 1,\ldots ,N \ , \ee
and any vector $|\Psi> \ \in {\mathcal{H}}_p$ can be expressed as a linear
combination of them. Especially each vector (70) is expressible as
\be
| \Psi^j> = \sum_{k=1}^N \xi_{kj} \ |  \chi_k> \ , \ j= 1,\ldots ,N \ee
and the constants
\be \xi_{kj} = <\chi_k| \Psi^j> \ . \ee

Evidently these relations could be expressed also in the form
\be
|\Psi^j> = \hat{\xi} | \chi_j>   \ , \ j=1,\ldots ,N  \ ,\ee
where
\be
\hat{\xi}    =    \sum_{k=1}^N  |\Psi^k><\chi_k|
    =   \sum_{\ell,k=1}^N \xi_{\ell k}|\chi_{\ell}><\chi_k| \ee
is a linear operator (on ${\mathcal{H}}_p$) and
\be
\xi_{kj} = < \chi_k| \hat{\xi}|\chi_j>   \ . \ee
Moreover, due to the linear independence of all $N$ vectors (70), the
relations (76) have to be invertible, i.e., the operator $\hat{\xi}$ is
nonsingular. Therefore the relations (76) are equivalently expressed as
\be
|\chi_j> = \hat{\eta} |\Psi^j> \ , \ j=1,\ldots ,N \ , \ee
where the linear operator (on ${\mathcal{H}}_p$)
\be
\hat{\eta} \equiv \hat{\xi}^{-1}  \ . \ee

Let us define  $ \forall \ |\Psi> \ \in \ {\mathcal{H}}_p$ the
corresponding vector (in ${\mathcal{H}}_p$)
\be
|\tilde{\Psi}> \equiv \hat{\eta} |\Psi> \ . \ee
Recalling Eq. (79) one can see that
\be
|\tilde{\Psi}^j> = | \chi_j> \ee
and hence the equivalence
\be
|\Psi> = \sum_{j=1}^N c_j | \Psi^j>  \ \Leftrightarrow |\tilde{\Psi}> =
\sum_{j=1}^N c_j | \chi_j> \ee
holds. Therefore, for any vector $|\Psi>, |\Phi> \ \in  {\mathcal{H}}_p$ the
relation
\be  \big(\Phi|\Psi> = <\tilde{\Phi}|\tilde{\Psi}>  \ee
holds. It means that the scalar product $\big(\cdot|\cdot>$ could be
expressed in terms of the scalar product $<\cdot|\cdot>$ as
\be
\big(\Phi|\Psi> = < \Phi |\hat{\kappa}|\Psi> \ , \ee
where the linear operator
\be
\hat{\kappa} \equiv \hat{\eta}^+\hat{\eta}  \ . \ee
It is useful to express Eq. (85) as a relation between two {\bf
different} bra-vectors related to the same ket by the two different
definitions of the scalar product (on the same {\bf vector} space
${\mathcal{H}}_p$) :
\be
\big( \Phi | = < \Phi | \hat{\kappa}  \ . \ee

One should not overlook that using this formalism one can e.g. express the
closure relation corresponding to the orthonormality condition (71) in the
following simple form
$$
\sum_{j=1}^N |\Psi^j>\big(\Psi^j| = 1  \ .\ \ \ \ \ \ \ \ \ \ \ \ \ {\rm 87a)}$$
Needless to say that this relation is equivalent to the closure
$$\sum_{j=1}^N |\chi_j><\chi_j| = 1 \ . $$

In order that the considered algorithm could provide a QT of a physical
system, one has to assign to each observable $A$ the corresponding
selfadjoint operator $\hat{A}$. At this point, it is important to keep in
mind that the notion of {\bf conjugation} of linear operators {\bf depends}
on the choice of the scalar product. We have tacitly used the symbol
$\hat{A}^+$ to denote the operator conjugated to the operator $\hat{A}$ in
the sense of the scalar product $<\cdot|\cdot>$, i.e., the relation
\be
<\Psi|\hat{A}^+|\Phi> = <\Phi|\hat{A}|\Psi>^* \ \ee
holds.

Let us introduce the symbol $\hat{A}^{\#}$ to denote the operator conjugated to
$\hat{A}$ in the sense of the scalar product $\big(\cdot|\cdot>$, i.e..  the
relation
\be
\big( \Psi|\hat{A}^{\#}|\Phi> = \big(\Phi | \hat{A} | \Psi>^* \ee
holds. By using the relation (87) one finds
\be
\big( \Phi| \hat{A} | \Psi>^*  =  < \Phi | \hat{\kappa} \hat{A} | \Psi>^*
 = < \Psi | \hat{A}^+ \hat{\kappa} | \Phi >
 =  \big( \Psi | \hat{\kappa}^{-1} \ \hat{A}^+ \ \hat{\kappa} | \Phi>  \ee
and therefore
\be
\hat{A}^{\#} = \hat{\kappa}^{-1} \ \hat{A}^+ \ \hat{\kappa}
= \hat{\xi}\hat{\xi}^+ \ \hat{A}^+ \ \hat{\eta}^+
\hat{\eta} \ .\ee
One can utilize the formal relations
\be
\big( |\Psi>\big)^+ = <\Psi|  \ \ ; \ \big( |\Psi>\big)^{\#} = \big(\Psi| \ ,
\ee
and therefore
\be
\big( |\Psi><\phi|\big)^+ = |\phi><\Psi| \ \  ; \
\big( |\Psi>\big(\phi|\big)^{\#} = |\phi>\big(\Psi|  \ , \ee
etc. It is also useful to notice that the operator (86) is selfadjoint
from the point of view of {\bf both} scalar products, i.e.
\be
\hat{\kappa} = \hat{\kappa}^+ = \hat{\kappa}^{\#} \ . \ee

Needless to say that when we required (below Eq. (87)) the
operator $\hat{A}$, assigned in the considered algorithm to an observable
$A$, to be selfadjoint, we meant that the relation
\be
\hat{A}^{\#} = \hat{A} \ee
has to hold.

The discussed algorithm may appear as a sort of QT which is still
``peculiar'' in two aspects:\\
i) the (total) energy is the only observable to which no selfadjoint
operator was assigned,\\
ii) evolution of the state has not been described by an ``ordinary
Schr\"odinger equation''.

Let us show now that both of these points could be circumvented, namely,
that the considered algorithm may be equivalently expressed in the form
of an ordinary quantum \\ mechanics:

Let us define the operator
\be
\hat{H} \equiv \sum_{j=1}^N z_j |\Psi^J>\big(\Psi^j | \ . \ee
From Eqs. (93), (54) and (70) follows that the
relations
\be
\hat{H}^{\#} = \hat{H}  \ , \ee
\be
\hat{H} |\Psi^j> = E_j |\Psi^j> \ , \
{\rm where} \
E_j  \equiv z_j \ , \ee
are guaranteed.
Moreover, from Eqs. (45), (59) and (70) follows that the solution
(43) with the initial condition
\be
|\Psi(t_0)> = |\Psi> \ \in \ {\mathcal{H}}_p \ee
may be expressed in the form
\be
|\Psi(t)> = \sum_j d_j e^{-iE_j t} |\Psi^j> \ , \
{\rm where} \
d_j \equiv \frac{c_j}{A^{(j)}} \ee
and hence the ``ordinary Schr\"odinger equation''
\be
i \frac{d}{dt} | \Psi(t)> = \hat{H} | \Psi(t) > \ee
holds.

Therefore the considered QT (with an ``energy dependent potential'') could
be equivalently rephrased as an ``ordinary quantum mechanics'' (QM1) in such
a way that\\
1) the Hilbert space of the corresponding system is identified with the
vector space ${\mathcal{H}}_p \ \subset \ \mathcal{H}$ endowed with the scalar
product $\big(\cdot|\cdot>$,\\
2) the Hamiltonian is identified with the operator (96).

Actually, one can rephrase the considered QT also as an ``ordinary quantum
mechanics'' (QM2) of the same physical system in such a way that the Hilbert
space assigned to this system is identified with the subspace
${\mathcal{H}}_p \ \subset \ \mathcal{H}$, i.e. with the vector space
${\mathcal{H}}_p $ endowed with the scalar product $<\cdot|\cdot>$.

To this end, it is sufficient to realize that once one associates to each
linear operator $\hat{A}$ the operator
\be
\hat{\tilde{A}} = \hat{\eta} \hat{A} \hat{\xi} \ , \ee
then, from the definition (81), one immediately sees that
$\forall |\Psi> \ \in \ {\mathcal{H}}_p$ \\
the relation
$|\Phi> = \hat{A} \ |\Psi> $
is equivalent to
$|\tilde{\Phi}> = \hat{\tilde{A}}  |\tilde{\Psi}> $, \\
and the relation
$
\hat{A}^{\#} = \hat{B}$
might be equivalently expressed as
$
\hat{\tilde{A}}^+ = \hat{\tilde{B}}$.

On the other hand, from the formulae (96), (87) and (82), one immediately
obtains

\be
\hat{\tilde{H}}  =   \sum_{j=1}^N z_j \hat{\eta}
|\Psi^j><\Psi^j|\hat{\eta}^+ \hat{\eta} \hat{\xi}
  =  \sum_{j=1}^N z_j |\chi_j><\chi_j | \  . \ee

In this case, the stationary states are described by the vectors
$|\chi_j>$
and the evolution of the states of the considered system is described by the
Schr\"odinger equation
\be
i \frac{d}{dt} | \tilde{\Psi}(t)> = \hat{\tilde{H}} | \tilde{\Psi}(t) > \ .\ee
The operators associated to observables are now selfadjoint in the sense of
the relation
\be
\hat{\tilde{A}}^+ = \hat{\tilde{A}} \ . \ee
We have seen that the considered algorithm could represent a QT if and only
if the operation $\big(\cdot|\cdot>$ (introduced as a part of this
algorithm) is expressible in the form (85), where $\hat{\kappa}$ is a
selfadjoint positive definite operator. How can one check whether this
condition is met? \\
First, it would be useful to express the operator $\hat{\eta}$ in
the form
\be
\hat{\eta} \equiv \sum_{j,k = 1}^N \eta_{jk} |\chi_j><\chi_k| \ .\ee
One finds that
\be
\eta_{jk} = <\chi_j|\hat{\eta}|\chi_k> = <\Psi^j|\hat{\eta}^+ \hat{\eta}|
\chi_k> = \big(\Psi^j|\chi_k> \ . \ee
This means that the coefficients on the r.h.s of Eq. (106) are identical with
the coefficients in the expression
\be
|\chi_j> = \sum_{k=1}^N \eta_{kj} | \Psi^k> \ .\ee
Therefore one can obtain the operator corresponding to any (arbitrarily
chosen) orthonormal basis (in the sense of relation (73))
\be
\big\{ \  |\chi_j> , \ j=1,\ldots ,N \big\} \ee
where each of its elements is expressed as a linear combination (108) with
the coefficients $\eta_{kj}$ utilized in Eq. (106).

Once the operator $\hat{\eta}$ is known, one can check whether all the
eigenvalues of the operator (94)
\be
\hat{\kappa} \equiv \hat{\eta}^+ \hat{\eta} \ee
are non-vanishing. (Needless to say that the operator $\hat{\kappa}$ is
uniquely determined by the operation $\big(\cdot|\cdot>$. The freedom in the
choice of the operator $\hat{\eta}$ is just a trivial consequence of the
fact that predictions of any (ordinary) quantum theory are invariant under
arbitrary unitary transformations.) If it is the case then the considered
algorithm could represent a QT. In the following we shall already tacitly
assume that the operation $\big(\cdot|\cdot>$ is introduced in the considered
algorithm in such a way that it can be interpreted as a scalar product in
${\mathcal{H}}_p$.

We have seen that this QT might be equivalently rephrased as a QM1 as well
as a QM2. For passing from the QT to the QM2, the knowledge of the operator
$\hat{\eta}$ appears essential (see Eqs. (102), (103) and (81)).
We have already discussed the steps which could in principle lead to the
construction of $\hat{\eta}$. Unfortunately, to proceed this way in practice
usually represents a formidably complicated task. One
naturally asks whether there exist any simpler way leading to the same goal.

In fact, it is easy to find a more straightforward procedure, at least
for one particular case. Let
\be
\hat{V}(z) = z\hat{V} \ee
and the operator $\big(\cdot|\cdot>$ is defined by the formulae (63)-(66). In
the case (111), the formula (64) simplifies to
\be
\hat{W}_{jk} = 1 - \hat{V} \ , \ee
and therefore (cf. Eqs. (63) and (85))
\be
\hat{\kappa} = 1 - \hat{V} \ . \ee
Such an algorithm might represent a QT
only if all values from the spectrum of the operator $\hat{V}$ are smaller
than 1. The evolution equation (43) can be rewritten as
\be
i \frac{d}{dt} (1 - \hat{V}) | \Psi(t)> = \hat{H}_0 |\Psi(t)> \ee
or equivalently as
\be
i \frac{d}{dt}| \tilde{\Psi}(t)> =  \hat{\xi}\hat{H}_0 \hat{\xi} |
\tilde{\Psi}(t)>  \ , \ee
where we have used the notation (81):
\be
|\tilde{\Psi}(t)> \equiv \hat{\eta} | \Psi(t)> \ , \ee
and without any loss of generality we have identified
\be
\hat{\eta} = \hat{\eta}^+ \equiv \sqrt{ \hat{\kappa}} = \sqrt{1 - \hat{V}} \
. \ee
The equation (115) is of course nothing else than the (ordinary)
Schr\"odinger eq. (104) with the Hamiltonian
\be
\hat{\tilde{H}} = \hat{\xi} \hat{H}_0 \hat{\xi} \ , \ee
where
\be
\hat{\xi} = [ 1 - \hat{V}]^{-1/2} \ .\ee
Therefore, in this case we have reached the QM2, which is equivalent to the
QT, just for free. Let us recall that in the framework of QM2, to each
observable $A$ is assigned an operator $\hat{\tilde{A}}$, which fulfills
the condition (105).
In particular (in the considered case), the operator assigned to the (total)
energy is given by formula (118).

From Eq.(102) follows that the operator $\hat{A}$, which is in
the framework of QM1 associated to the observable $A$, is related to the
operator $\hat{\tilde{A}}$ associated to the same observable in the
framework of QM2 as follows
$$
\hat{A} = \hat{\xi} \hat{\tilde{A}} \hat{\eta} \ ,  $$
i.e.
\be
\hat{A} = \frac{1}{\sqrt{(1 - \hat{V})}} \hat{\tilde{A}} \sqrt{(1 - \hat{V})}
\ . \ee
Especially, the Hamiltonian in the QM1 is the operator
\be
\hat{H} = \hat{\xi}\hat{\tilde{H}}\hat{\eta} = \hat{\xi}^2\hat{H}_0 \ , \ee
i.e.
\be \hat{H} = \frac{1}{1 - \hat{V}} \hat{H}_0 \ .\ee

\section{ The toy model revisited.}

In the light of the preceding section, we are going to re-examine our
toy model. We will present an example of QT which
reduces to QM1 or QM2. Let us
consider the special case when
\be
{\mathcal{H}} \equiv L^2(-\infty,\infty) \ , \ee
\be \hat{H}_0 \equiv - \frac{1}{2} \frac{d^2}{dx^2} + V_0(x) \ , \ee
\be
\hat{V} = V(x) \ , \ee
and the real function $V$ fulfills the condition
$$
V(x) < 1 \ \ \forall x \ \in \ (-\infty,\infty) \ . $$
Physically, this should be a one dimensional version of QT of a particle in
an external field.

In this case
\be
\hat{\tilde{H}} = \frac{1}{ 1 - V}
\Big\{
V_0 - \frac{1}{2 } \Big[
\frac{d^2}{dx^2} + \frac{V'}{1 - V} \ \frac{d}{dx} + \frac{1}{4} \
\frac{2V''(1 - V) + 3 V'^2}{(1 - V)^2} \Big]  \Big\}  \ . \ee
To be more specific, let us take
\be
V(x) = - K x^2 \ , \ K>0 \ . \ee
Then
\be
\hat{\tilde{H}} = \frac{1}{1 + Kx^2} \Big\{ V_0(x) - \frac{1}{2 } \Big[
\frac{d^2}{dx^2} - \frac{2Kx}{1 + Kx^2} \ \frac{d}{dx} - K \
\frac{1 - 2Kx^2}{(1 + Kx^2)^2} \Big]  \Big\}  \ . \ee
To obtain the energy spectrum, one has to find those values of the parameter
$E$ for which the differential equation
\be
\Big\{ - \frac{1}{2 } \Big[
\frac{d^2}{dx^2} - \frac{2Kx}{1 + Kx^2} \ \frac{d}{dx} - K \
\frac{1 - 2Kx^2}{(1 + Kx^2)^2} \Big] + V_0(x) \Big\} \chi_E(x) = E(1 +
Kx^2)\chi_E(x)  \ee
has solutions in ${\mathcal{H}}_p$ of QM2, i.e., which satisfy the
integrability condition
\be
\int_{-\infty}^{\infty} |\chi_E(x)|^2 dx < \infty  \ . \ee
(For simplicity sake, we are explicitly describing here the procedure
which concerns only the discrete part of the energy spectrum).

To be even more specific, let us consider the special case when
\be
V_0(x) = Ax^2 \ , \ A > 0 \ .\ee
Then Eq. (129) yields
\be
\frac{1}{2 } \chi''  + \frac{Kx}{(1 + Kx^2)} \chi' - \Big[ \frac{K}{2 } \
\frac{1 - 2Kx^2}{(1 + Kx^2)^2} + E  + (EK - A) x^2 \Big] \chi = 0 \ . \ee
In the asymptotic region $|x| \rightarrow \infty$, Eq. (137) acquires the form
\be
\chi'' + \lambda^2 x^2 \chi \simeq 0 \ , \ee
where
\be
\lambda^2(E) = 2 (A - KE) \ . \ee
Thus, for each real value of the parameter E, one can choose 2
independent solutions of Eq. (132) in such a way that their behavior
in the asymptotic region  $x \rightarrow \infty$ is dominated by the factor
\be
e^{-\frac{\lambda(E)}{2}x^2} \ \ {\rm and} \ \ e^{\frac{\lambda(E)}{2}x^2} \
, \ee
respectively. Similarly, one can choose 2 independent solutions of (132)
in such a way that their behavior in the asymptotic region
$x \rightarrow -\infty$ is also dominated by the factor (135).

For
$
E > \frac{A}{K}$,
$\lambda(E)$ is purely imaginary, and consequently both independent solutions
oscillate in both asymptotic regions. This is why we strongly suspect that
each value
$
E \ \in (\frac{A}{K}, \infty)$
represents a (twice degenerated) value from the continuous part of the
spectrum of the operator
$\hat{\tilde{H}}$.
We are not going to dwell on closer
examination of this point here.

Let us turn to the examination of Eq. (132) for
$
E < \frac{A}{K}$.
Due to the invariance of the operator $\hat{\tilde{H}}$ under the replacement
$
x \ \rightarrow \ -x $,
for each value of the parameter $E < A/K$,
there could be at most one solution of Eq. (132) belonging
to $L^2 (-\infty, \infty)$. This means that the whole discrete part of the
spectrum is non-degenerated. To find the explicit formula for this part
of the energy spectrum, let us express the function $\chi_E(x)$ as
\be
\chi(x) = \sqrt{\lambda + Ky^2} f(y) e^{-y^2/2} \ , \ {\rm where } \
y = \sqrt{\lambda} x  \ . \ee
Then  Eq. (132) implies that the function $f$ has to solve the equation
\be
f'' - 2 y f' + ( \frac{2E}{\lambda} - 1)f = 0 \ . \ee
It is well known from elementary QM that\\
i) if \be  \frac{2E}{\lambda} = 2n + 1 \ , \ee
where $n$ is an even (odd) nonnegative integer, then
in the asymptotic region $|y| \rightarrow \infty$,
symmetric
(antisymmetric) solutions of Eq. (137)
\be
f(y) \propto H_n(y) \ , \ee
where $H_n$ is the Hermite polynomial, whereas antisymmetric (symmetric)
solutions behave as
\be
f(y) \sim e^{y^2} \ . \ee
\\
ii) If the condition (138) is not satisfied, then the general solution of Eq.
(137) has the asymptotic behavior (140).

Therefore, the discrete part of the energy spectrum is indeed non-degenerate
and its energy levels are determined by Eqs. (138) and (134) as
\be
E_n = \frac{2n + 1}{4 } \Big[ \sqrt{8A + (2n+1)^2K^2 } - (2n+1)K \Big] \ .
\ee
One should not overlook that $E_n$ is a monotonically increasing function of
$n$ with the upper limit
\be
\lim_{n\to\infty} E_n = \frac{A}{K} \ . \ee

The stationary state with the energy (141) is described in the framework of
QM2 by the wave function
\be
\chi_n(x) = C_n \sqrt{1 + Kx^2} H_n(\sqrt{\lambda_n}x)
e^{-\lambda_n x^2/2}\
,\ee
where $C_n$ is a normalization constant and
\be
\lambda_n \equiv 2 \sqrt{A - KE_n} \ .\ee

In the framework of QM1, the same stationary state is described by the wave
function
\be
\psi^n(x) \equiv \hat{\xi} \chi_n(x) = C_n H_n(\sqrt{\lambda_n}x)
e^{-\lambda_n x^2/2} \ . \ee

The same wave function describes this state also in the QT.

The above analysis sheds  some  light on the toy models of section 2.
In particular, the linear $E$-dependence with $\gamma < 0 $ coincides with
the example presented in this chapter by setting $A = 1/2$ and $K = |\gamma|$.
In this special case (as one can see from formulae (92), (118)) the ansatz (9)
looks like relation (87a) expressed in the "x-representation". It means that this
ansatz would indeed be valid provided that\\
a) the ``physical Hilbert space" ${\mathcal{H}}_p $ is isomorphic to $L^2(-\infty,\infty)$\\
b) there is no continuous part of the energy spectrum.\\
As we have already stressed these conditions are apparently not met. Therefore, even in this
special case the r.h.s. of the formula (9) is only an approximation of its l.h.s.

Positive values of $\gamma$ are unsuitable. According to Eq (117), the
operator $\hat{\eta}$ reduces to zero at some place, becomes complex and is
not invertible, no matter how small is $\gamma$. This is the origin of the
questionable results concerning the sign of $<x^k>_n$ or the overestimate of
$<x^2>_0$ and $<x^4>_0$ when applying the closure relation (Tables 1 and 2).
The situation is similar for the $\sqrt{E_n}$-dependence as regards to the
sign of $\gamma$. However,
constructing the corresponding $\hat{\eta}$ and $\hat{\kappa}$ operators
can be quite tedious. It may also happen that a judicious choice of the
$|\chi>$ vectors suggests a useful approximation. It lies outside the
scope of the present work to investigate this point further.

Let us now return to the more general case of the one dimensional QT
determined by Eqs. (123)-(125):
the Hamiltonian (126) can be expressed as
\be
\hat{\tilde{H}} = \hat{\tilde{T}} +  \hat{\tilde{H}}_{(I)} \ ,\ee
where the first term
\be
\hat{\tilde{T}} \equiv \frac{ \hat{\tilde{P}}^2}{2} \ee
describes the kinetic energy of the considered particle.
\be
\hat{\tilde{H}}_{(I)} \equiv \frac{1}{1 - V} \Big\{ V_0 - \frac{1}{2} \Big[
\frac{2V'' (1 - V) + 3 V'^{2}}{ 4(1 - V)^2} + i \frac{V'}{1 - V}
\hat{\tilde{P}} - V \hat{\tilde{P}}^2 \Big] \Big\} \ee
describes its interaction with the external field and
\be
\hat{\tilde{P}} \equiv - i \frac{d}{dx} \ee
is the operator assigned to its momentum in the framework of QM2.

It should be noticed that\\
i)  rephrasing the QT (with energy dependent potential) as QM2 (ordinary
quantum mechanics) revealed that the interaction of the considered
particle is {\bf momentum} dependent.\\
ii) The expression for the Hamiltonian $\hat{H}$ describing the considered
physical system in the framework of QM1 is obtained from Eqs. (146)
- (149) simply by omitting all {\it tildes} in these expressions.\\
iii) In the framework of QM1 the momentum is described by the operator (cf.
(102))
\be
\hat{P} \equiv \hat{\xi} \hat{\tilde{P}} \hat{\eta} = \hat{\tilde{P}} +
\hat{\xi} \big[ \hat{P},\hat{\eta} \big] \ , \ee
i.e.
\be
\hat{P} = -i \frac{d}{dx} + \frac{i}{2} \frac{V'(x)}{ \sqrt{1 - V(x)}}
\ . \ee
iv) If one insists to work in the framework of QT with an energy dependent potential,
one has to identify again the momentum operator with the operator (151).

\section{Conclusions.}

The present work is devoted to wave equations with energy dependent
potentials. Calculations performed in the framework of several toy models
 clearly demonstrated that "plausible" modification of the "scalar product"
 does not guarantee intrinsic consistency of the resulting "quantum theory
 with an energy dependent potential".

 In order to show what sort of caution one has to observe towards such straightforward
 implementation of energy dependent potentials in quantum theory we have resorted to a more
 fundamental analysis of this problem. The main moral following from the analysis could be
 summarized as follows:\\
 i) Proper care has to be paid to the assignment of Hilbert space to the considered physical
 system. Especially one has to observe that\\
 a) vectors corresponding to stationary states with different energies have to be
 orthogonal \footnote{One should keep in mind that in the framework of "theories with energy 
dependent potentials" (in contrast to "ordinary" quantum theories) this is not guaranteed for 
free. As a direct consequence, the specification of the energy dependent potential is not 
sufficient for  determination of the dynamics of the considered physical system (and of the 
energy spectrum, as well) predicted by such theory.}\\
b) closure relation expressed in terms of stationary state wave functions properly reflects their 
normalization\\
c) operators corresponding to observables are all indeed selfadjoint. \\
ii) Once a "quantum mechanics with an energy dependent potential" does not suffer
by intrinsic inconsistencies one can rephrase it into a form of ordinary quantum mechanics.
The price one has to be ready to pay for such a transition to "ordinary theory" is a
possible (additional) momentum dependence at that part of the Hamiltonian
which should describe interaction.

\bigskip
{\bf Acknowledgments}

We would like to express our thanks to H. Sazdjian for valuable discussions
and comments.
This work was supported by the agreement between IN2P3 and ASCR
(collaboration no. 97-13) and by the Grant Agency of the Czech
Republic (J.M., grant No. IAA1048305).\\
J.F. acknowledges the hospitality of the Group of Theoretical Physics,
IPN Orsay.

\begin{center}
\begin{figure}
\epsfig{file=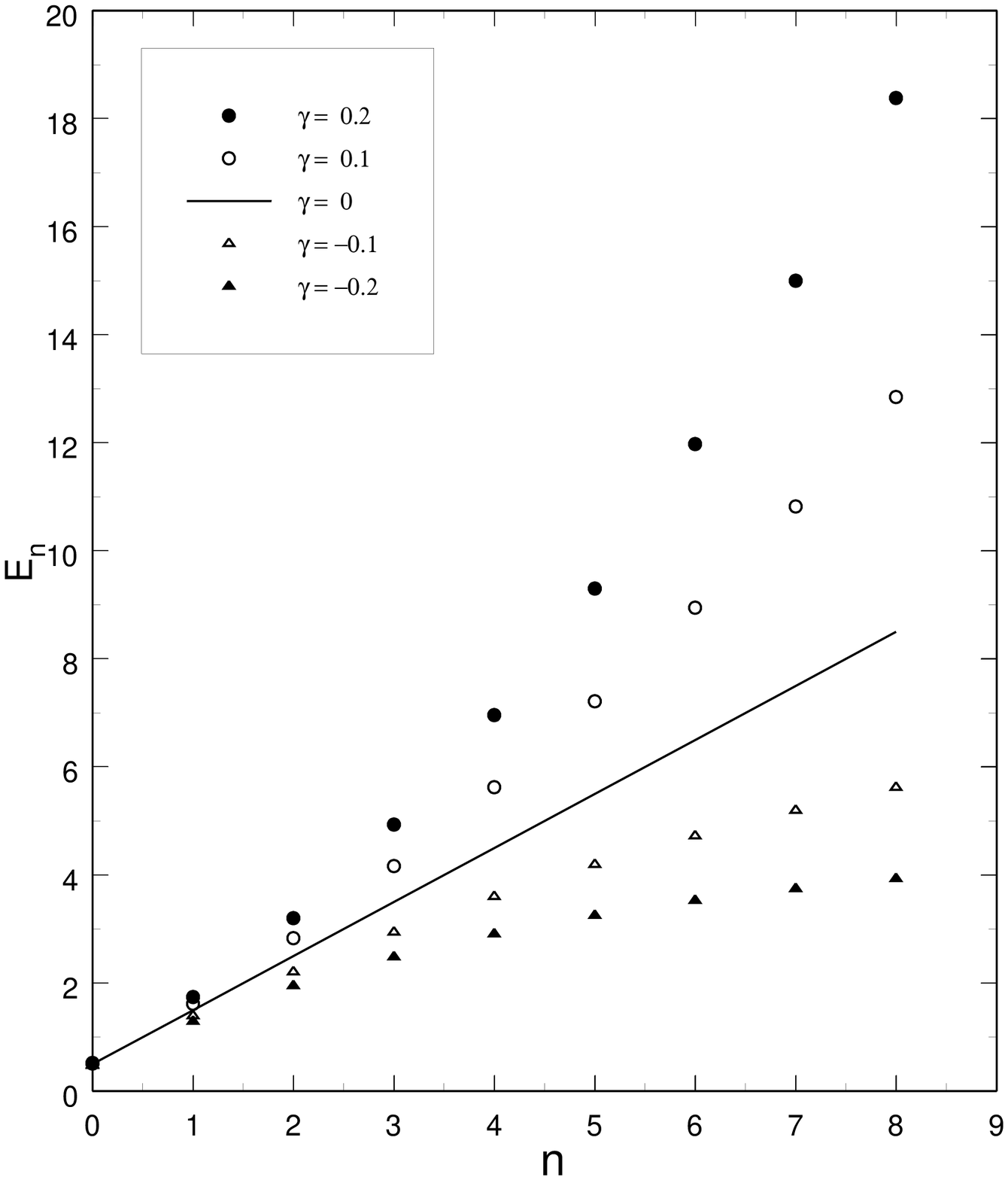, width=15cm}
\caption{
$E_n$ as function of n for the linear E-dependent harmonic
oscillator, with $\gamma=0, \pm 0.1, \ {\rm and} \  \pm 0.2$. The solid line
corresponds to $\gamma=0$, points above (below) correspond
to $\gamma>0$ ($\gamma<0$).
}
\end{figure}
\end{center}

\newpage
\begin{table}
\caption
{Comparison of $<x^2>_0$ as function of $\gamma$
for linear E dependence with predictions of the linear approximation
(3rd column) and the sum (28); $n_{max}$ denotes the highest
state involved in the sum. See text for details.
}
\vspace*{3mm}
\begin{center}
\begin{tabular}{|c|c|c|c|c|}\hline
  &  &  &  &  \\
$\gamma$ & $<x^2>_0$ & $\frac{1}{2}(1 - \frac{3}{4}\gamma)$  & $n_{max} = 1$ &
$n_{max} = 3$ \\
  &  &  &  & \\\hline
  & &  &  &  \\
 0.01 &     0.496258 &     0.496250  &    0.496253 &     0.496258\\
0.05  &    0.481445 &     0.481250  &    0.481338  &    0.481445\\
0.10 &     0.463281  &    0.462500  &    0.462894  &    0.463275\\
0.25  &    0.411124  &    0.406250  &    0.409449  &    0.411012\\
0.50  &    0.331895  &    0.312500  &    0.329503  &    0.332217\\
  &  &  &  &   \\\hline
  & &  &  &  \\
- 0.01 &     0.503758  &    0.503750  &    0.503753 &     0.503758\\
- 0.05 &     0.518945  &    0.518750  &    0.518818 &     0.518945\\
- 0.10 &     0.538281  &    0.537500  &    0.537731 &     0.538270\\
- 0.25 &     0.598624  &    0.593750  &    0.594457 &     0.598111\\
- 0.50 &     0.706895  &    0.687500  &    0.686108 &     0.699289\\
  &  &  &  &  \\\hline
\end{tabular}
\end{center}
\end{table}

\begin{table}
\caption{Comparison of $<x^4>_0$ as function of $\gamma$
for linear E dependence with predictions of the linear approximation
(3rd column) and the sum (30). See text for details.
}
\vspace*{3mm}
\begin{center}
\begin{tabular}{|c|c|c|c|c|c|}\hline
  &  &  &  &  & \\
$\gamma$ & $<x^4>_0$ & $\frac{3}{4}(1 - \frac{3}{2}\gamma)$
& $n_{max} = 2$& $n_{max}=4$ & $n_{max} = 6$ \\
 &  &  &  &  & \\\hline
  & & &  &  &  \\
0.01 &  0.738797 &  0.738750 &   0.738760 &   0.738797 &   0.738797\\
0.05 &  0.694912 &  0.693750 &   0.694091 &   0.694904 &   0.694911\\
0.10 &  0.642105 &  0.637500 &   0.639288 &   0.642006 &   0.642097\\
0.25 &  0.496767 &  0.468750 &   0.487138 &   0.495611 &   0.496552\\
0.50 &  0.294496 &  0.187500 &   0.292966 &   0.301638 &   0.302978\\
 &  &  &  & & \\\hline
  & & &  &  &  \\
-0.01 &   0.761297 &   0.761250 &   0.761259 &   0.761297 &   0.761297\\
-0.05 &   0.807432 &   0.806250 &   0.806378 &   0.807421 &   0.807432\\
-0.10 &   0.867270 &   0.862500 &   0.862603 &   0.867066 &   0.867248\\
-0.25 &   1.061827 &   1.031250 &   1.025211 &   1.053279 &   1.057984\\
-0.50 &   1.439879 &   1.312500 &   1.263341 &   1.340323 &   1.361776\\
 &  &  &  & & \\\hline
\end{tabular}
\end{center}
\end{table}

\begin{table}
\caption
{Test of the dipole sum rule for linear $E$ dependence and $\gamma>0$
for several maximum values $n_{max}$ in the sum (32). The exact value is
given by the last term in Eq. (32).
}
\vspace*{3mm}
\begin{center}
\begin{tabular}{|c|c|c|c|c|c|}\hline
 &  &  &  &  & \\
 $\gamma$ & $n_{max} = 1$ & $n_{max} = 3$ &$n_{max} = 5$&$n_{max} = 7$
 & exact\\
   & &  &  &  &  \\\hline
  &  &  &  &  & \\
 0.01  &  0.501236  &  0.501250 &   0.501250 &   0.501250  &  0.501250\\
 0.05  &  0.505894  &  0.506247 &   0.506249 &   0.506250  &  0.506250\\
 0.10  &  0.511061  &  0.512451 &   0.512492 &   0.512495  &  0.512496\\
 0.25  &  0.522114  &  0.529647 &   0.530583 &   0.530834  &  0.531189\\
 0.50  &  0.526582  &  0.545979 &   0.550165 &   0.551709  &  0.562017\\
  & &  &  &  & \\\hline
  &  &  &  &  &   \\
-0.01 &   0.498736 &   0.498750 &   0.498750  &  0.498750 &   0.498750\\
-0.05 &   0.493404 &   0.493748 &   0.493750  &  0.493750 &   0.493750\\
-0.10 &   0.486139 &   0.487461 &   0.487500  &  0.487503 &   0.487504\\
-0.25 &   0.460802 &   0.467450 &   0.468276  &  0.468497 &   0.468811\\
-0.50 &   0.410368 &   0.425484 &   0.428746  &  0.429950 &   0.437983\\
  & &  &  &  & \\\hline
\end{tabular}
\end{center}
\end{table}

\begin{table}
\caption
{Test of the "modified" closure relation (10). $\Psi_2$
and $\Psi_4$ denotes the 1st and the 2nd contribution to
the sum (10), respectively.
}
\vspace*{3mm}
\begin{center}
\begin{tabular}{|c|c|c|c|c|}\hline
  &  \multicolumn{2}{c|}{}  & \multicolumn{2}{c|}{}   \\
   & \multicolumn{2}{c|}{$\gamma > 0$} &
\multicolumn{2}{c|}{$\gamma < 0$} \\
   &  \multicolumn{2}{c|}{}  & \multicolumn{2}{c|}{}   \\\hline
   &  &  &  & \\
 $|\gamma|$ &  $\Psi_2$ & $\Psi_4$ & $\Psi_2$ & $\Psi_4$\\
   & &  &  &   \\\hline
  &  &  &  &  \\
0.01 &    -0.000428  &    0.000000  &    0.000426  &    0.000000\\
0.05 &    -0.002122  &   -0.000030  &    0.002145  &   -0.000033\\
0.10 &    -0.004211  &   -0.000116  &    0.004301  &   -0.000139\\
0.25 &    -0.010212  &   -0.000628  &    0.010670  &   -0.000954\\
0.50 &    -0.019131  &   -0.001965  &    0.019540  &   -0.003851\\
  &  &  &  & \\\hline
\end{tabular}
\end{center}
\end{table}

\begin{table}
\caption
{$<x^2>_0$ as function of $\gamma$, $\gamma>0$,
for $\sqrt{E}$ dependence. Exact values (exact) are compared with the
sums $\sum^{n_{max}}_n |<n|x|0>|^2$, $n_{max}$ = 1,3 and 5.
See text for details.
}
\vspace*{3mm}
\begin{center}
\begin{tabular}{|c|c|c|c|c|}\hline
  &  &  &  &  \\
$\gamma$ & $n_{max}=1$ & $n_{max}=3$ & $n_{max}=5$ & exact\\
  &  &  &  &  \\\hline
  & &  &   &  \\
0.01  &      0.498530  &    0.498534  &  0.498534 & 0.498229\\
0.05  &      0.484265   &   0.484270 &   0.484280 & 0.482782\\
0.10  &      0.469378   &   0.469396  &  0.469434 & 0.466445\\
0.25  &      0.429326   &   0.429411  &  0.429591 &0.422212\\
0.50  &      0.375328   &   0.375554  &  0.376051 & 0.361802\\
  &  &  &  &  \\\hline
  & &  &  &  \\
-0.01 &     0.501465 &     0.501469   &  0.501469 &   0.501765\\
-0.05 &     0.516501 &     0.516507   &  0.516521 &   0.518156\\
-0.10 &     0.533654 &     0.533682   &  0.533745 &   0.537310\\
-0.25 &     0.588962 &     0.589213   &  0.589857 &   0.601416\\
-0.50 &     0.691896 &     0.693731   &  0.702316 &   0.734997\\
  &  &  &   &  \\\hline
\end{tabular}
\end{center}
\end{table}
\end{document}